\documentclass[amsfonts, amssymb, amsmath, reprint, showkeys, twoside, prl, floatfix]{revtex4-2}
\usepackage[english]{babel}

\usepackage[utf8]{inputenc}
\usepackage[T1]{fontenc}

\usepackage{amsthm}
\usepackage{mathtools}
\usepackage{units}

\usepackage{xcolor}
\usepackage[pdftex]{graphicx}

\usepackage{hyperref}
\hypersetup{colorlinks=true, citecolor=blue, urlcolor=blue, linkcolor=blue}

\usepackage{multirow}

\usepackage[textsize=tiny,tickmarkheight=1ex]{todonotes}

\newcommand{\bra}[1]{\langle#1\rvert} 
\newcommand{\ket}[1]{\lvert#1\rangle} 
\newcommand{\expect}[1]{ \langle #1 \rangle} 

\newcommand{\operator}[1]{\hat{#1}}

\newcommand{\cone}{\mathrm{i}}

\renewcommand{\vec}[1]{\boldsymbol{#1}}

\begin{document}
\title{Ultrafast Dynamics of Orbital Angular Momentum of Electrons Induced by Femtosecond Laser Pulses: Generation and Transfer Across Interfaces}

\author{Oliver Busch}
\email[Correspondence email address: ]{oliver.busch@physik.uni-halle.de}
\author{Franziska Ziolkowski}
\author{Ingrid Mertig}
\author{Jürgen Henk}
\affiliation{Institut für Physik, Martin Luther University Halle-Wittenberg, 06099 Halle, Germany}

\date{\today}

\begin{abstract}
The orbital angular momenta (OAM) of electrons play an increasingly important role in ultrafast electron and magnetization dynamics. In this theoretical study, we investigate the electron dynamics induced by femtosecond laser pulses in a normal metal, a ferromagnet, and a ferromagnet/normal metal heterostructure. We analyze the spatio-temporal distributions of the laser-induced OAM and their respective currents. Our findings demonstrate that a circularly polarized laser pulse can induce a sizable and long-lasting OAM component in a normal metal. Furthermore, an interface between a ferromagnet and a normal metal facilitates the demagnetization of the magnet by the OAM contribution to the total magnetization. Finally, to transfer OAM from a ferromagnet into a normal metal, it is advantageous to use a laser setup that induces the desired OAM component in the ferromagnet, but not in the normal metal.
\end{abstract}


\maketitle

\paragraph{Introduction.} In recent years, there has been significant attention given to ultrafast phenomena in condensed matter physics. While much of the focus has been on the spin angular momentum (SAM) of electrons, which has led to the field of spintronics, the closely related orbital angular momentum (OAM) of electrons has also emerged as an important topic in its own right~\cite{boeglin2010, stamm2010, hennecke2019}. Orbitronic devices are seen as a potential alternative to electronic and spintronic devices~\cite{bernevig2005, go2021, go2020}.

A number of ultrafast phenomena are well described by means of SAM, for example, the optical manipulation of magnetic moments~\cite{duong2004, kimel2007, kirilyuk2010}, the demagnetization of ferromagnets~\cite{beaurepaire1996,zhang2000,zhang2003} and the transfer of magnetic moment between ferromagnetic layers~\cite{malinowski2008,schellekens2014,rudolf2012ultrafast} as well as across magnet/normal metal interfaces~\cite{chen2019,melnikov2011ultrafast,alekhin2017femtosecond}. Moreover, femtosecond laser pulses induce SAM in non-magnetic and magnetic samples~\cite{Neufeld2023,busch2023ultrafast}.

Spin-orbit coupling is not only ubiquitous in solids but also indispensable for most of the phenomena mentioned above. Therefore, a question arises regarding the contributions of the OAM to these effects (recall that SAM and OAM add up to the total angular momentum). In this respect, we need to address several issues, such as: what components of the OAM are induced by femtosecond laser pulses, what is their magnitude, and what is their spatio-temporal distribution? In this Paper, we report on a theoretical study where we investigated photo-induced OAM and their currents in Cu(100), Co(100), and a Co/Cu(100) heterostructure excited with femtosecond laser pulses.

Our findings, based on investigations of a Cu(100) film, reveal a pronounced and persistent presence of laser-induced OAM in the direction of the OAM current which propagates through the sample. Furthermore, in the context of a Co/Cu(100) heterostructure, we observe that the interaction at the interface between a ferromagnet and a normal metal facilitates the demagnetization of the magnet due to the OAM contribution to the overall magnetization. Notably, for efficient transfer of OAM from the ferromagnet to the normal metal, careful consideration should be given to the polarization configuration of the laser pulse. Specifically, inducing the desired OAM component exclusively within the ferromagnet, rather than in the normal metal, proves advantageous.

Our study offers valuable insights into the ultrafast dynamics of electron orbital angular momenta. These dynamics are determined by both the electronic and magnetic properties of the samples as well as by the laser pulse characteristics.

\begin{figure}
    \centering
    \includegraphics[width=0.9\columnwidth]{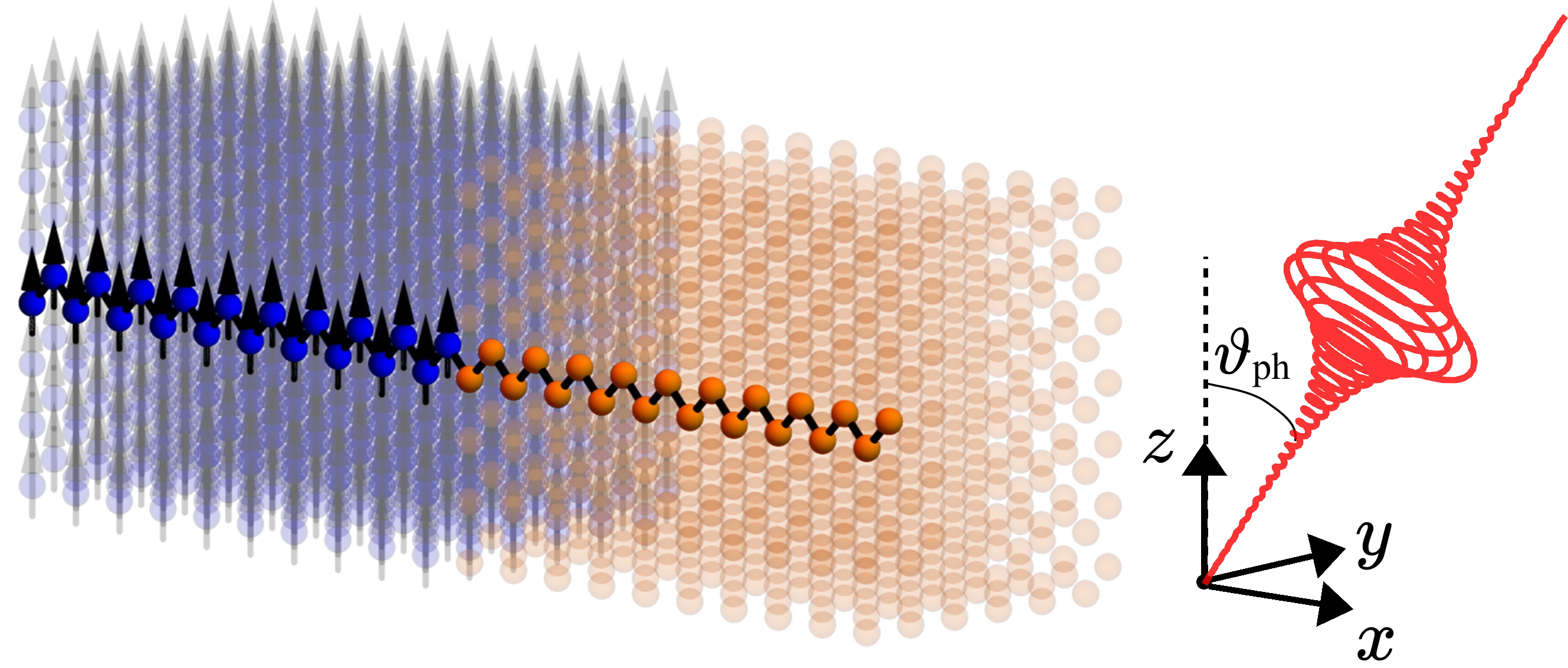}
    \caption{Geometry of an fcc Co/Cu(100) heterostructure. The film is composed of 40 layers stacked in the $x$ direction, with 20 layers of both Co atoms (blue spheres) and Cu atoms (orange spheres). It is infinite in both the $y$ and the $z$ direction. The Co magnetic moments point along the $z$ direction (black arrows). A circularly polarized laser pulse is incident within the $xz$ plane onto the sample.}
    \label{fig:sketch_Co_Cu}
\end{figure}

\paragraph{Theoretical aspects.} We briefly present the main ideas of our approach to ultrafast electron dynamics, \textsc{evolve}, since it has been described elsewhere~\cite{Toepler2021, busch2023ultrafast}.

We consider free-standing films of Cu(100), face-centered cubic Co(100), and Co/Cu(100), each with a thickness of 40 layers (20 layers each for Co/Cu). The Cartesian $x$-axis is perpendicular to the film, and periodic boundary conditions are applied in the $y$ and $z$ directions. The local magnetic moments in Co(100) and Co/Cu(100) are collinear and point along the $z$ direction (Fig.~\ref{fig:sketch_Co_Cu})~\cite{Heinrich91}.

The electron dynamics is described by the von Neumann equation (in Hartree atomic units)
\begin{align}
   -\cone \frac{\mathrm{d} \operator{\rho}(t)}{\mathrm{d} t} & =  [ \operator{\rho}(t),\operator{H}(t)]
    \label{eq:EOM}
\end{align}
for the one-particle density matrix
\begin{align}
    \operator{\rho}(t) = \sum_{n, m} \ket{n} \, p_{nm}(t) \, \bra{m}.
\end{align}
$\{ \ket{n} \}$ are the eigenstates of the Hamiltonian $\operator{H}_0$ which describes the electronic structure of the samples in tight-binding form~\cite{Slater1954,Papaconstantopoulos2015}. Collinear magnetism and spin-orbit coupling are included~\cite{Konschuh2010}. 

The time-dependent Hamiltonian $\operator{H}(t)$ in~\eqref{eq:EOM} supplements $\operator{H}_0$ by the electric field of the femtosecond laser pulse ~\cite{Savasta1995}. This field is a coherent superposition of s- and p-polarized partial waves with energy $\omega$ and with a Lorentzian envelope. In this Paper we focus on excitation by circularly polarized light with helicity $\sigma_{+}$ impinging within the $xz$~plane onto the films with a polar angle $\vartheta_{\mathrm{ph}} = 45^{\circ}$ of incidence (Fig.~\ref{fig:sketch_Co_Cu}).

The geometry of the entire setup dictates what components of the orbital angular momentum $\expect{\vec{L}}$ can be produced by the incident radiation~\cite{Henk1996,busch2023ultrafast}. As for the spin angular momentum, all three components $\expect{L^{\mu}}$ ($\mu = x, y, z$) can be induced in both non-magnetic and magnetic samples by circularly polarized light.

The spatio-temporal properties of an observable $O$ are obtained by taking partial traces in the expectation value $\expect{O}(t) = \operatorname{tr}[\operator{\rho}(t) \,\operator{O}]$, with the density matrix in an appropriate basis; partial trace means that the trace is restricted to the desired subspace, e.g., to a specific site, orbital or OAM component. We address the OAM $\expect{\vec{l}_{i}}(t)$ at site $i$ and its site-averaged (global) companion
\begin{align}
    \expect{\vec{L}}(t) & \equiv \frac{1}{N} \sum_{i} \expect{\vec{l}_{i}}(t)
    \label{eq:orbpol-average}
\end{align}
in which the summation is over the $N$ sites in a sample's unit cell. Similarly to the SAM currents in Ref.~\onlinecite{busch2023ultrafast}, OAM currents are computed from the symmetrized form
\begin{align}
	\expect{j_{kl}^{\mu}}(t) & \equiv \frac{1}{2} \left( \expect{L^{\mu} j_{kl}}(t) + \expect{j_{kl} L^{\mu}}(t) \right), \quad \mu = x, y, z,
\end{align}
in which the operator $\operator{\jmath}_{kl}$ for the current from site $l$ to site $k$ is derived from Mahan's expression~\cite{mahan2013,busch2023ultrafast}. Here, we focus on non-equilibrium currents across the films (along the zigzag path in Fig.~\ref{fig:sketch_Co_Cu}), since these are important for OAM transfer within stacked samples (as often used in experiments).

In all simulations discussed below, the laser has a photon energy of~$\unit[1.55]{eV}$, a fluence of about $\unit[3.3]{mJ\,cm^{-2}}$, and is modulated with a Lorentzian with a width of $\unit[10]{fs}$ and centered at $t = \unit[0]{fs}$. All samples comprise $40$~layers, with sites~$0$ and~$39$ defining the bottom and top surfaces, respectively.

\begin{figure}
    \centering
    \includegraphics[width=0.8\columnwidth]{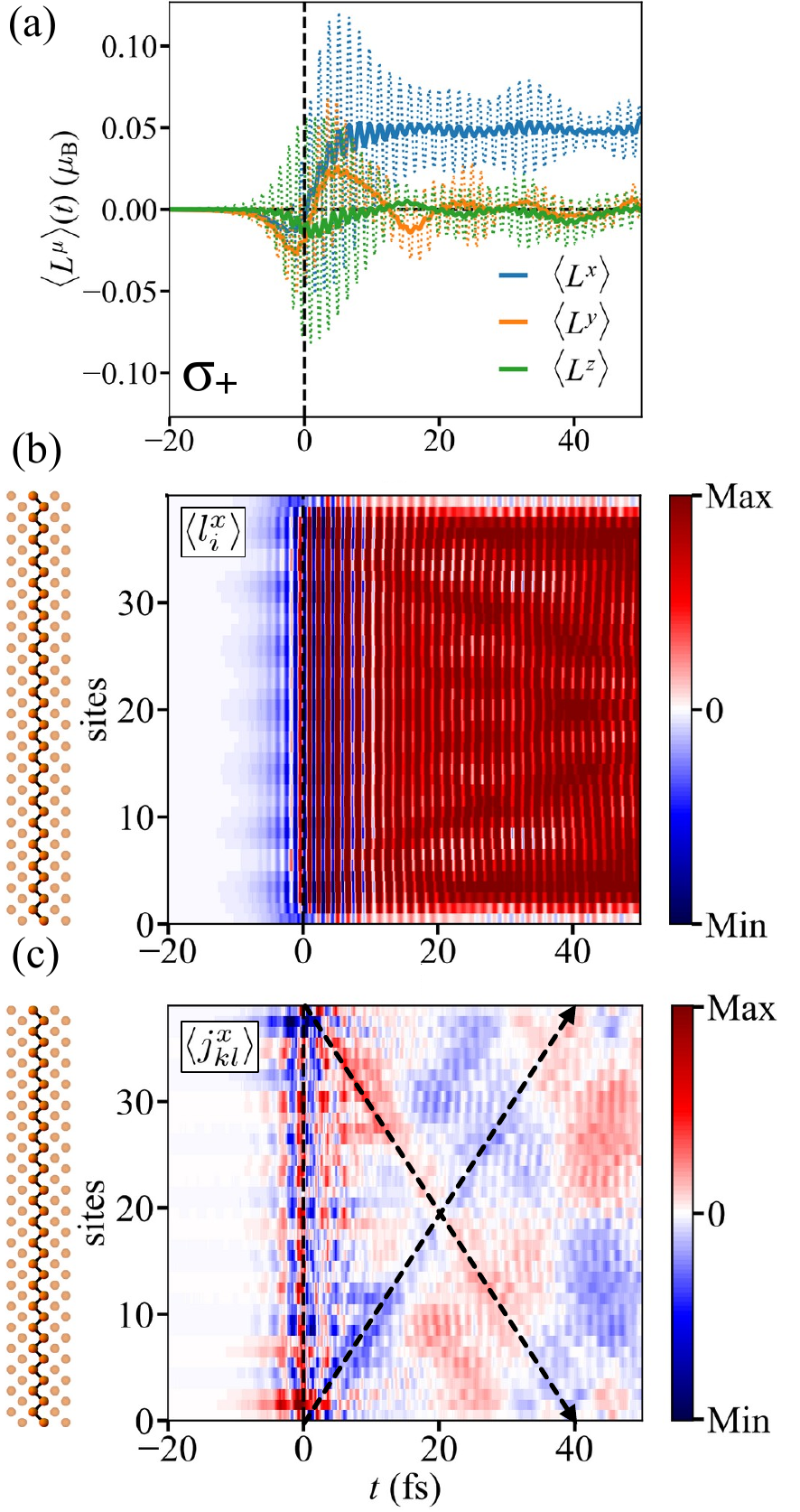}
    \caption{Photo-induced orbital angular momentum (OAM) in Cu(100) is excited with a circularly polarized laser pulse with helicity $\sigma_{+}$. (a) Site-averaged (global) OAM $\expect{L^{\mu}}(t)$, $\mu = x, y, z$. Thicker lines represent the data convoluted with a Gaussian with standard deviation $\sigma = 10$ to better visualize the main trends. (b) Spatio-temporal distribution of $\expect{l^{x}_{i}}(t)$ depicted as a color scale. (c) OAM current $\expect{j_{kl}^{x}}(t)$ across the sample. Arrows indicate the criss-cross pattern. The color bars indicate positive (red) and negative values (blue) of $\expect{l^{x}}$ and $\expect{j_{kl}^{x}}$ in panels (b) and (c), respectively. Dashed vertical lines at $t = \unit[0]{fs}$ mark the laser-pulse maximum.}
    \label{fig:Cu_circpol}
\end{figure}

\paragraph{Cu(100).} A circularly polarized laser pulse induces all three OAM components [panel~(a) of Fig.~\ref{fig:Cu_circpol}]. As has been found for the SAM~\cite{busch2023ultrafast}, all components of $\expect{\vec{L}}$ exhibit rapid oscillations that are associated with the laser's frequency. Both $\expect{L^{y}}$ and $\expect{L^{z}}$ fluctuate slowly about $\unit[0]{\mu_{\mathrm{B}}}$ after the pulse. Strikingly, $\expect{L^{x}}$ is increased within $\unit[10]{fs}$ and oscillates about an almost constant value of $\unit[0.05]{\mu_{\mathrm{B}}}$ per site [thick blue spectrum in panel~(a)]. This finding implies that an OAM component of measurable magnitude persists considerably long after the femtosecond laser pulse in a non-magnetic sample.

The spatial distribution of $\expect{l_{i}^{x}}$ is uniform across the sample, as evidenced by the negative (blue) onset before the laser-pulse maximum and the plateau-like positive (red) distribution after the laser pulse [panel~(b)]. Minor deviations from this uniformity result in an OAM current $\expect{j_{kl}^{x}}$, which is strongest at the surfaces [sites~0 and~39; panel~(c)] at $t\approx \unit[0]{fs}$ and moves toward the center of the sample. This creates an antisymmetric criss-cross pattern [arrows in panel~(c)]. In a semi-infinite sample, an $x$-polarized OAM current $\expect{j_{kl}^{x}}$ would be observed starting at the surface, similar to the OAM of electron beams, which is also oriented in the propagation direction~\cite{Grillo2017}.

\begin{figure}
    \centering
    \includegraphics[width=0.8\columnwidth]{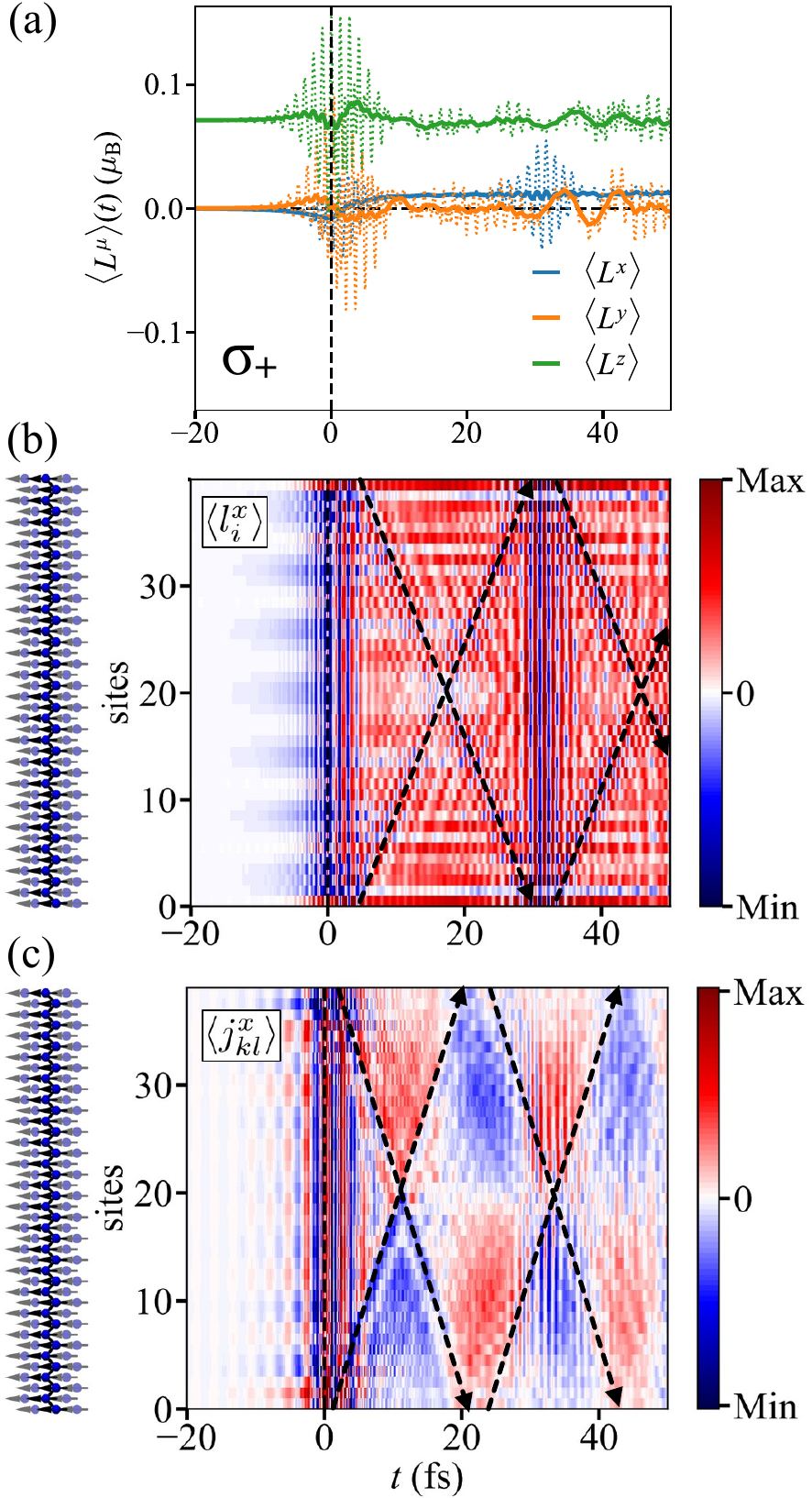}
    \caption{Same as Fig.~\ref{fig:Cu_circpol} but for face-centered-cubic Co(100). Recall that the OAM of Co has an intrinsic $z$ component of $\unit[0.07]{\mu_{\mathrm{B}}}$ [green in panel~(a)].}
    \label{fig:Co_circpol}
\end{figure}

\paragraph{Co(100) and Co/Cu(100).} After successfully establishing sizable and long-lasting laser-induced OAM in copper, we now shift our attention towards magnetic systems.

For a Co(100) sample, one can observe a reduction of the SAM, well-known as demagnetization~\cite{busch2023ultrafast,beaurepaire1996}. The $z$-component of the OAM is strongly modulated during the pulse but remains constant thereafter [shown in green in panel (a) of Fig.~\ref{fig:Co_circpol}] and has a magnitude similar to the intrinsic $\expect{l^{z}}$ ($\unit[0.07]{\mu_{\mathrm{B}}}$, which is in agreement with published data~\cite{Igarashi1996}). This result suggests that OAM does not significantly contribute to demagnetization in homogeneous magnetic samples, at least in fcc Co. It is worth noting that a laser pulse with opposite helicity, $\sigma_{-}$, also does not lead to orbital demagnetization (not shown here).

\begin{figure*}
    \centering
    \includegraphics[width=0.9\textwidth]{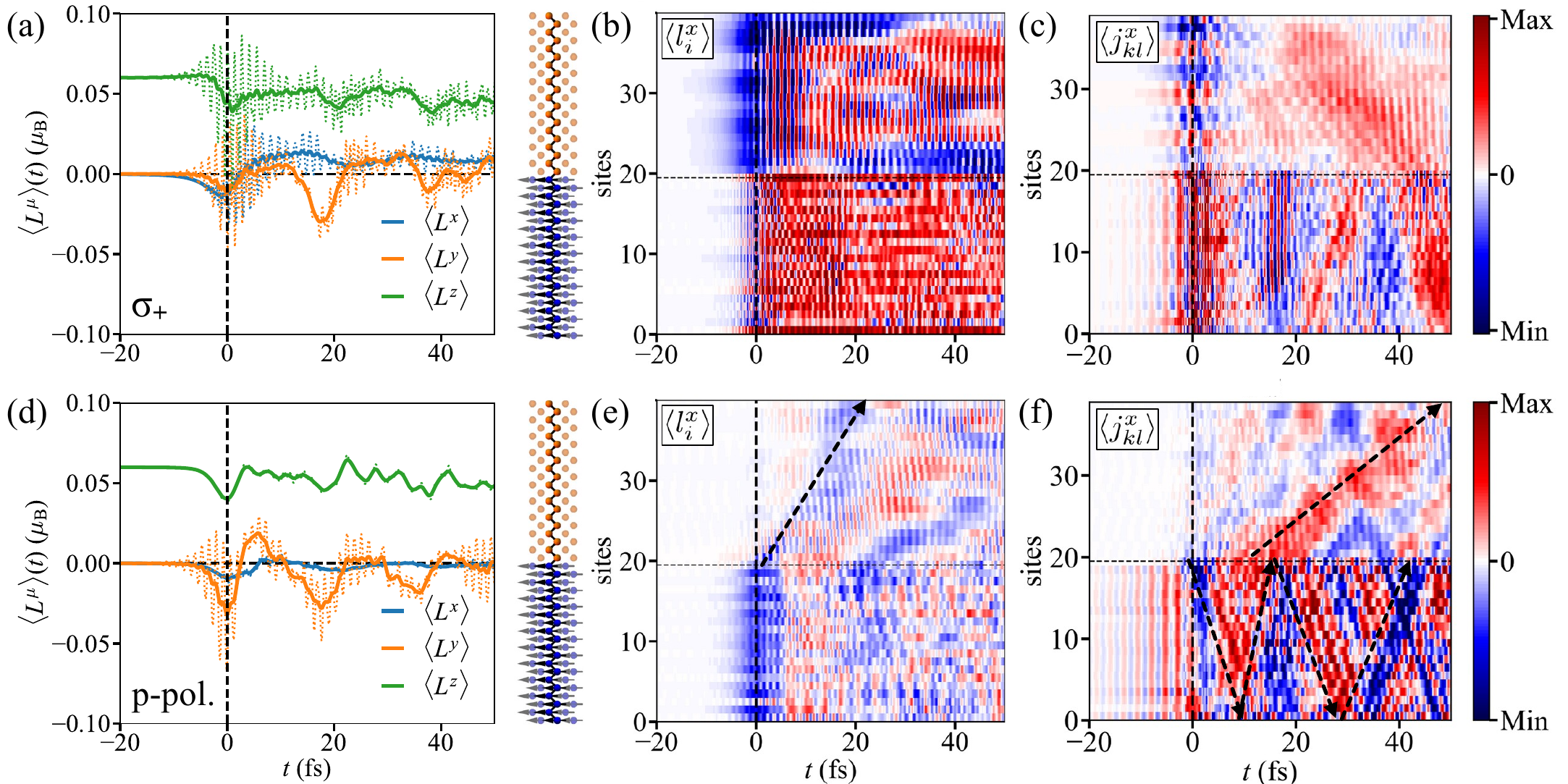}
    \caption{Same as Fig.~\ref{fig:Cu_circpol} but for a Co/Cu(100) heterostructure excited with circularly polarized [top row, panels~(a)--(c)] and p-polarized laser pulse [bottom row, panels~(d)--(f)].}
    \label{fig:CoCu_Lx_p_and_circ_pol}
\end{figure*}

In the context of the Co/Cu(100) heterostructure, $\expect{L^{z}}$ is reduced to approximately $\unit[0.04]{\mu_{\mathrm{B}}}$, representing a relative decrease of roughly $\unit[35]{\%}$ [green in panel (a) of Fig.~\ref{fig:CoCu_Lx_p_and_circ_pol}]. This finding suggests that inhomogeneities play an important role in orbital demagnetization, a phenomenon that has already been established for the SAM\@. In an inhomogeneous sample, the interface acts as a source for both SAM and OAM currents, which we attribute to the local imbalance of spin-dependent (SAM) or $\expect{l_{i}^{z}}$ occupation.

In contrast to Cu(100), $\expect{L^{x}}$ is relatively small in both Co(100) and Co/Cu(110) (about $\unit[0.01]{\mu_{\mathrm{B}}}$) but persists, as well. Moreover, there is no precession of $\expect{\vec{L}}$ before the pulse maximum \footnote{A simple explanation is that the Hamiltonian $\operator{H}_{0}$ does not contain OAM operators but SAM operators in order to account for exchange splitting (Zeeman term). The Heisenberg equations of motion for the angular-momenta components thus yield precession for the SAM but not for the OAM\@.}, which differs from the SAM~\cite{busch2023ultrafast}.

Inspecting the spatio-temporal OAM-current distributions shows a criss-cross pattern for Co, similar to that for Cu. The pattern for Co/Cu(100) is slightly more complicated, but neither indicates a pronounced transfer of OAM across the interface, especially from Co into Cu.

For Co/Cu(100) we found a pronounced dependence of the SAM distribution on the laser's polarization, which suggests to replace circularly polarized light by p-polarized light (electric field oscillates in the $xz$ plane; Fig.~\ref{fig:sketch_Co_Cu}) in order to evoke transfer of OAM from Co into Cu. Recall that $\expect{L^{x}}$ is not induced by p-polarized light in Cu but in Co~\cite{busch2023ultrafast} \footnote{The symmetry analysis for the SAM, reported in Ref.~\onlinecite{busch2023ultrafast}, also holds for the OAM\@.}.

Indeed, the negative $\expect{l_{i}^{x}}(t)$ induced at about $t = \unit[0]{fs}$ in the Co region [dark blue region for sites~0 to~19; panel~(e) of Fig.~\ref{fig:CoCu_Lx_p_and_circ_pol}] is transferred into the Cu region (oblique blue stripe starting at the interface). In addition, the oscillations of $\expect{l_{i}^{x}}$ shortly after the laser pulse (red-blue from $\unit[10]{fs}$ to $\unit[20]{fs}$) propagate into the Cu region, visible as oblique stripes. The pattern is perhaps better visible in the distribution of $\expect{j_{kl}^{x}}$ [panel~(f)]: the criss-cross motif discussed before `spills over' from the Co region into the Cu region. This holds for the $z$-OAM as well (not shown here). Consequently, it is possible to transfer OAM from a ferromagnet into a normal metal using an appropriate laser pulse.

\paragraph{Concluding remarks.} Our theoretical investigation yields these results: a sizable and persistent OAM component can be induced in Cu by a circularly polarized laser pulse; an interface between a ferromagnet and a normal metal facilitates the demagnetization of the magnet, not only for the SAM but also for the OAM contribution to the total magnetization; in order to transfer OAM from a ferromagnet across an interface into a normal metal it appears advantageous to use a setup in which the respective OAM component is \emph{not} induced in the normal metal; this concerns, in particular, the polarization of the femtosecond laser pulse.

Standing to reason, these findings call for experimental verification, which might be challenging. It is not just that experiments on ultrafast timescales are demanding, it may be intricate to disentangle the spin and orbital contributions to the total angular momentum. As suggested in Ref.~\onlinecite{go2018}, a suitable method for probing orbital currents could be similar to the indirect detection of spin currents via accumulated angular momentum at the edges of a sample via the magneto-optical Kerr effect (MOKE)~\cite{kato2004, sinova2006, stamm2017}. In addition, X-ray magnetic circular dichroism (XMCD) measurements~\cite{obrien1994, bonetti2017} allow to discriminate SAM and OAM; being element-specific they provide also details on the OAM in regions of heterostructures. Moreover, we consider it worthy to investigate other materials and material combinations.

This work is funded by the Deutsche Forschungsgemeinschaft (DFG, German Research Foundation) -- Project-ID 328545488 -- TRR~227, project~B04.

\bibliographystyle{apsrev4-2}
\bibliography{references}

\end{document}